\documentclass[twocolumn,superscriptaddress,longbibliography,aps,pra,preprintnumbers]{revtex4-1}

\usepackage{graphicx}
\usepackage{bm}
\usepackage{color}
\usepackage{epstopdf}
\usepackage{amsmath}
\usepackage{amssymb}
\usepackage{epstopdf}
\usepackage{lipsum}
\usepackage{psfrag}

\usepackage[urlcolor=blue,colorlinks=true,citecolor=blue,linkcolor=blue,pdfstartview={FitH},bookmarks=false]{hyperref}

\newcommand{\ket}[1]{|{#1}\rangle}

\newcommand{\expectation}[3]{\langle{#1}|{#2}|{#3}\rangle}

\definecolor{olive}{rgb}{0., 0.5, 0.}

\graphicspath{{fig/}{./fig/}{.}}

\begin{document}

%\preprint{To be submitted to: Physical Review A}

\title{Retrieval of free-Majorana wavefunctions for the finite Kitaev chain using an appropriate momentum representation}
%  Momentum space and topology in the 1D Kitaev lattice systems

\author{Karen Rodr\'{i}guez}
\email[e-mail: ]{karem.c.rodriguez@correounivalle.edu.co}

\author{Ang\'elica Alejandra P\'erez-Losada}
\email[e-mail: ]{angelica.perez@correounivalle.edu.co }
\affiliation{Departamento de F\'{i}sica, Universidad del Valle, A.A. 25360, Cali, Colombia}
\affiliation{Centre for Bioinformatics and Photonics -- CiBioFi, Calle 13 No. 100-00, Edificio 320 No. 1069, Cali, Colombia}

\author{Arturo Arg\"uelles}
\email[e-mail: ]{arturo.arguelles00@usc.edu.co}
\affiliation{Universidad Santiago de Cali, Facultad de Ciencias B\'asicas, Campus Pampalinda, Calle 5 No. 62-00, C\'odigo postal 76001, Santiago de Cali, Colombia}

\date{\today}% It is always \today, today,

\begin{abstract}
  In quantum mechanics, the spaces of momentum and its conjugate, the position, are related via Fourier transforms and thus the properties are interwoven with their structure. In particular, for lattice systems possessing an underlying discrete position space, the momentum becomes finite. Moreover, if the lattice length is finite, $L<\infty$, the momentum space also becomes both finite and discrete breaking altogether the continuity of the dispersion relation. This aspect is relevant in new systems such as the topological materials. We address this point paving the path for new ways to the observation of Majorana quasi-particles. Furthermore, the Kitaev model, which is the simplest Hamiltonian supporting Majorana fermions, is therefore taken as a starting point for the theoretical description of the work. Our study focuses on finding the zero-energy modes in an artificial arrangement of a non-interacting superconducting finite nanowire by using a discrete-sine transform with the purpose of going from position to momentum space considering hard-wall boundary conditions in the process.  Here, a new $L$-dimensional Nambu operator to diagonalize the system in the Bogoliubov-de Gennes formalism is proposed and, further, we arrive to a new space with a considerable reduced dimension allowing the treatment of larger system sizes.
 % where the Majorana representation is diagonal.
  We also present a comparison between the numerical and third order perturbation theory for the weak coupling regime results presenting an excellent agreement. Finally, we analyze the wavefunctions of the retrieved zero-energy modes showing that they are indeed edge states and thus they have an exponential decay.
\end{abstract}


\maketitle

\section{Introduction}

Nature always surprises us with novel wonders. Nowadays, the condensed matter community focuses efforts on  rare materials such as topological insulators that behave as insulators in the inside keeping conducting states on its surface~\cite{PhysRevLett.95.146802}. Although band insulators may present surface states with conducting properties, the main feature of the topological version is that they present a  non-trivial symmetry-protected topological order~\cite{PhysRev.56.317,PhysRevB.85.075125}. This fact can be used to achieve nanometric scale holograms using nanometric topological insulator thin films as an optical resonant cavity~\cite{Yue2017}, to name but an instance.

On the other hand, one of the most attractive phenomena expected to appear in high-energy physics is unveiling itself as low energy excitations in condensed matter systems. This is the particular case of both Weyl and Majorana fermions~\cite{PhysRevLett.107.127205,Ciudad2015,Wilczek2009}. The former appear as solution of the Dirac equation for massless spin-1/2 particles~\cite{Weyl1929} and were discovered as collective excitations in TaSa since its linearly dispersing valence and conduction bands cross at discrete points enabling the material to host them~\cite{Xu613}. More recently, new effects have come to light as the Kondo effect on such materials rendering them as Weyl-Kondo semimetals~\cite{Lai93}, widening the possibilities of innovation and technology out of this field. The later rise from superconducting topological materials where the surface states may host non-fundamental excitations known as Bogoliubov quasiparticles that are their own antiparticles thus bringing them the nickname of Majorana zero modes. 

The Majorana fermions idea permeates several fields going from high energy physics theories~\cite{Avignone2008,Frank2009}, to condenced matter experimental realizations~\cite{Mourik2012,Williams2012,Deng2012,Anindya2012} including quantum computing and spintronics applications. In the later, the Majoranas are presented in form of non-trivial emergent excitations rather than fundamental particles.

The immense development of experimental techniques in ultracold atomic Fermi gases in the last years has opened new avenues for the research of strongly correlated systems in condensed matter physics and beyond. The ability to control the interactions via Feshbach resonances~\cite{fedichev.prl.96} sets new perspectives for experimental realization and study of several different exotic systems such as spin-polarized superfluidity (with population imbalance), superconductivity with nontrivial Cooper pairing and Bose-Fermi mixtures or mixtures of fermions with unequal masses, among others~\cite{zwierlein.schirotzek.06, partridge.li.06,georgescu.ashhab.14}. 
Currently, the unconventional superconductivity with a non-trivial Cooper pairing lays down one of the most important directions of studies in the theory of condensed matter~\cite{matsuda.shimahara.10} and ultracold quantum gases~\cite{hu.xiaji.06,dutta.gajda.15}. 

A remarkable simple model holding unpaired Majorana fermions has been presented by Kitaev~\cite{Kitaev2001}, which is a model of topological superconductor comprised of spinless fermions and thus is the model we consider. It can be experimentally realized by a semiconducting finite nanowire device able to accommodate Cooper pairs through an artificial arrangement holding a strong spin-orbit interaction in proximity to an s-wave superconductor under the action of a external magnetic field~\cite{Mourik2012,Williams2012,Deng2012,Anindya2012}. Moreover, the Kitaev chain is equivalent to the Su-Schrieffer-Heeger (SSH) model~\cite{PhysRevB.96.165124} that has several experimental realizations where the topological features are observed~\cite{Meier2016,Belopolskie1501692}.

Keeping all that in mind, this paper is organized as follows. In Sec.~\ref{sec.2} we briefly introduce the Kitaev model and use the Bogoulibov-de Gennes formalism in momentum space to retrieve its energy spectrum however missing the unpaired Majoranas due to the unbroken translational symmetry. Then, we employ a revised version of the Fourier transform that imposes hard-wall boundary conditions allowing the rise of edge states. Later, in Sec.~\ref{Perturbation}, we consider a weak coupling regime where the pairing is treated as a small perturbation up to 3rd order on a tight-binding model to retrieve the system spectrum. Then, Sec.~\ref{Wavefunction} focuses on the edge states retrieving the wave function in both momentum and position spaces showing, in the later, their localization and exponential decay. And finally, Sec.~\ref{Conclusions} is devoted to conclusions and remarks.

\section{Model and method}

\label{sec.2}
\begin{figure}[t!]
\includegraphics[width=0.45\textwidth]{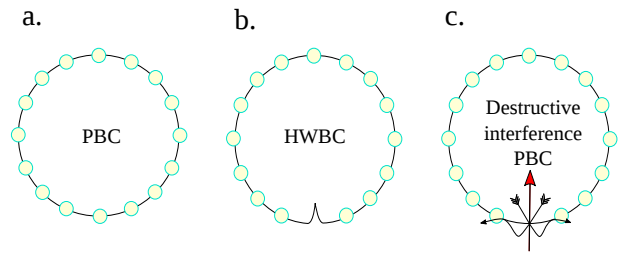}
\caption{\label{PBC_HWBC} One dimensional chain depicted with a) periodic boundary conditions (PBC) and b) hard-wall boundary conditions (HWBC). Panel c) presents the special PBC case for $k=\{0,\pi\}$ where destructive interference appear at a given lattice point recovering the topology of the HWBC.}
\end{figure}

We consider the Kitaev model described by the Hamiltonian~\cite{Kitaev2001}
\begin{equation}
\label{Kitaev-position}
\hat H=\sum_l\left[\left(-t \hat a_l^\dagger\hat a_{l+1}-\Delta\hat a_l\hat a_{l+1}+\text{h.c.}\right)-\mu\left(\hat a_l^\dagger\hat a_l-\frac{1}{2}\right)\right],
\end{equation}
where $\hat a_l^\dagger$ ($\hat a_l$) is the operator that creates (annihilates) a fermion at the $l$-th site, the hopping rate is given by $t$, $\mu$ is the chemical potential and $\Delta$ is the pairing parameter or superconducting gap.  
\paragraph{Translational invariant momentum representation:} The customary Fourier transform relating momentum and position spaces, $\hat a_l=(1/\sqrt{L})\sum_k \exp(-ikl)\hat a_k$ with $l$ the lattice site index and $k$ the momentum index, holds only for periodic boundary conditions with a periodicity of $L$ and thus it may even represent infinite chains in the uniform case. It is worth to mention that there exist two particular points, $k=\{0,\pi\}$, where  destructive interference breaks the translational invariance as shown in Fig.~\ref{PBC_HWBC}. Hence, only in that case the system behaves as a finite chain. %Nevertheless, this fact although suitable for non-interacting systems is useless for the more realistic case of interacting particles.

 %----------------------DIFFERENCE BETWEEN BOUNDARY CONDITIONS-------------
%-------------------------------------------------------------------------

%----------------KITAEV AND FOURIER TRANSFORM-------------------------
\begin{figure}[t!]
\resizebox{8.3cm}{!}{\input{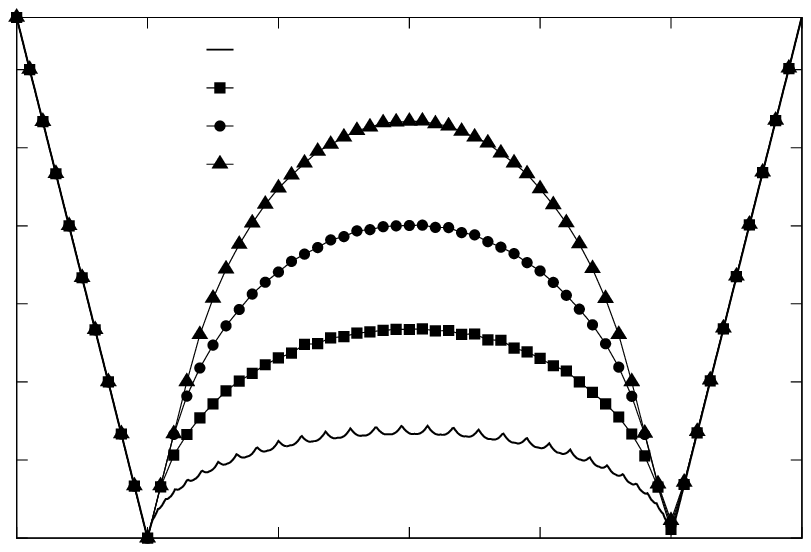}}
\caption{\label{PBC_Kitaev}The Kitaev chain described by the usual Fourier transform (periodic boundary conditions), leads to wrong gap results since it does not close where the topological phase is expected within $\mu\in[-2,2]$. The plots where generated by taking $L=51$.}% \Ka{Esta gr\'afica falta arreglarle el tipo de l\'inea y el tama\~no de la fuente del key}}
\end{figure}
%----------------------------------------------------------------------

The Hamiltonian (\ref{Kitaev-position}) represented in momentum space is written in a matrix form by using the Nambu operators $\hat C_k=(\hat a_k, \hat a^\dagger_{-k})^\textrm{T}$ and $\hat C_k^\dagger=(\hat a^\dagger_k, \hat a_{-k})^\textrm{T}$. Following the Bogoliubov and de Gennes (BdG) formalism~\cite{BdG1999}, it takes the form $\hat H=\sum_k \hat C^\dagger\mathbb H_{\text{BdG}}\hat C+\sum_k\epsilon_k$ where the coupling matrix is given by,
\begin{equation}
\mathbb H_{\text{BdG}}=\begin{pmatrix} \epsilon_k&-\Delta_k^*\\-\Delta_k&-\epsilon_k\end{pmatrix}
\end{equation}
and $\epsilon_k=-\mu-2t\cos(k)$, $\Delta_k=i\Delta\sin(k)$. % with $\Delta$ a real number.
The eigenvalues of $\mathbb H_{\text{BdG}}$ are: $\lambda_{\pm}=\mp\sqrt{\epsilon_k^2+|\Delta_k|^2}$. In Fig.~\ref{PBC_Kitaev}, we plot the difference $\Delta E=|\lambda_+-\lambda_-|$ for a set of $\Delta\in(0,1)$, being $t$ the energy unit, i.e. $t\equiv 1$. For \mbox{$\Delta>0$}, when the gap is finite the phase is trivial, otherwise the system has a topological phase supporting free Majorana states~\cite{Alicea2012}. As it is observed, the gap closes only at $|\mu|=2$ and for $\Delta\ll 1$, the number of visible wrinkles is the exactly $L$ (see curve $\Delta=0.2$ in Fig.~\ref{PBC_Kitaev}). Hence, the formalism is not suitable to describe the system with unpaired Majoranas since the expected results are not obtained.

   %Figure~\ref{PBC_Kitaev} shows that the Kitaev chain supporting free Majoranas at the edges is not properly described in this picture, since the gap does not close in the chemical potential window $\mu\in[-2,2]$.

  \paragraph{Sine discrete Fourier transform type I:}
  
  In order to seek the unpaired Majoranas, we should express the Kitaev model in the so-called Majorana representation in position space. The usual relations are $\hat a_l=\frac{1}{2}\left(\hat \gamma_l^A+i\hat\gamma_l^B\right)$, and $\hat a_l^\dagger=\tfrac{1}{2}\left(\hat \gamma_l^A-i\hat\gamma_l^B\right)$, where $\hat \gamma_l^\alpha$ creates/annihilates a Majorana quasi-particle of the $\alpha\in\{A,B\}$ kind at the $l$-th site. These operators hold the anticommutation relation $\{\hat\gamma_l^A,\hat\gamma_m^B\}=\delta_{l,m}\delta_{A,B}$ and $\left(\hat\gamma_l^\alpha\right)^2=1$, which leads the quasi-particles to be their own antiparticles or, in other words, $\hat\gamma^A=(\hat\gamma^A)^\dagger$. In this way, the Hamiltonian (\ref{Kitaev-position}) in the Majoranas representation reads

  \begin{align}\label{Kitaev-Majoranas}
\hat H&=-i\sum_j\left[(t+\Delta)\hat\gamma_j^A\hat\gamma_{j+1}^B+(t-\Delta)\hat\gamma_{j+1}^A\hat\gamma_j^B\right]\notag\\
&~~~-i\mu\sum_j\hat\gamma_j^A\hat\gamma_j^B .
\end{align}

As we have seen previously, the typical Fourier transform leaves out the possibility of having edge states, therefore, we use an specific transform that complies with the required boundary conditions. To achieve this, we consider the kinetic eigenfunctions given by plane waves, where due to the degeneracy,  every pair of states with opposite momenta $\lbrace|\pm k\rangle\rbrace$ form a subspace, and on that account, one can use a convenient basis given by $\ket{\zeta^{\beta}}=\tfrac{e^{\frac{i \pi}{4}(1-\beta)}}{\sqrt{2}}\left[\ket{k}+\beta\ket{-k}\right]$ with $\beta=\pm 1.$ In order to be consistent with the hard-wall boundary conditions the edge states require, we impose the eigenfunctions to vanish at the edges ($l=0$ and $l=L+1$) leaving us only with the states $\ket{\zeta^{-}}$ and an effective system size of $L$. The basis orthonormality is granted since $\langle \zeta^-|\zeta'^-\rangle=\delta_{\zeta,\zeta'}$ (see appendix~\ref{OSDFT}). The corresponding sine discrete Fourier transform type I on the ladder operators is finally given by
\begin{equation}
\label{sine-transform}
\hat a_l=\sqrt{\frac{2}{L+1}}\sum_{\zeta=1}^{L}\sin\left(\frac{\pi l \zeta}{L+1}\right)\hat a_\zeta. 
\end{equation}

%-----------------USUAL KITAEV PHASE DIAGRAM---------------------------------
\begin{figure}[t!]
\centering
\vspace{-1cm}
\resizebox{8.3cm}{!}{\input{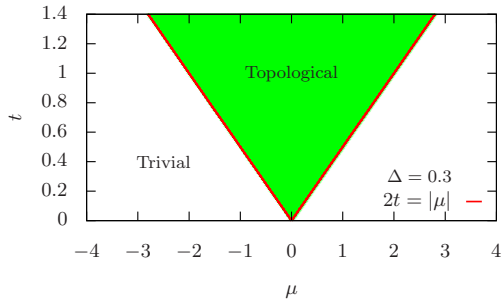}}
\caption{\label{Kitaev_usual-PD} (Color online.) Usual phase diagram for the Kitaev chain, the topological phase is located in the color region bounded by the $2t=|\mu|$ lines, outside the system remains in the so-called trivial phase.}
\end{figure}
%-------------------------------------------------------------------------

After applying this appropriate transform to equation (\ref{Kitaev-Majoranas}) we are led into the momentum representation where the Hamiltonian takes the form
\begin{equation}
\label{H-Majoranas}
\hat H=i\sum_{\zeta=1}^L\hat\gamma_\zeta^A\left[E_\zeta-2\Delta\sum_{\zeta'}F_{\zeta\zeta'}\right]\hat\gamma_{\zeta'}^B,
\end{equation}
with
\begin{align}
E_\zeta(\mu)&=-\mu-2t\cos\left(\frac{\pi\zeta}{L+1}\right),\notag\\
F_{\zeta\zeta'}&= \frac{2}{L+1}\left(\frac{\textrm{S}_\zeta\textrm{S}_{\zeta'}}{\textrm{C}_\zeta-\textrm{C}_{\zeta'}}\right),~\zeta+\zeta'\text{ odd},
\end{align}
where $\textrm{S}_\zeta=\sin\left(\frac{\pi\zeta}{L+1}\right)$ and $\textrm{C}_\zeta=\cos\left(\frac{\pi\zeta}{L+1}\right)$. Accordingly, the numerical problem has been reduced from an exponentially large matrix dimension of ($2^L\times 2^L$) to ($L\times L$), allowing us to study much larger system sizes without losing any information during the process since no approximation has been done.

The notation is significantly simplified by the use of the following  Nambu-like operators: \hbox{$\vec\Gamma^\alpha=(\hat\gamma_1^\alpha,\dots,\hat\gamma_\zeta^\alpha,\dots,\hat\gamma_L^\alpha)^\text{T}$}, where $\alpha\in\{A,B\}$. In this way, the Hamiltonian (\ref{H-Majoranas}) becomes,
\begin{equation}\label{BdG}
\hat H=i\vec\Gamma^A\cdot\left[\mathbb H_\text{D}+\mathbb H_\text{S}\right]\vec\Gamma^B,
\end{equation}
with $\mathbb H_\text{D}$ being a diagonal matrix displaying the dispersion relation of free fermions in the chain and $\mathbb H_\text{S}$ is a skew-symmetric matrix containing the pairing term that switches on the topological phase along with the free Majoranas. Figure~\ref{Kitaev_usual-PD} presents the usual Kitaev's phase diagram for which the color region holds the topological phase whereas in the white region the phase remains trivial. The figure is obtained by calculating the singular value decomposition (SVD) of the Bogoliubov-de Gennes coupling matrix, $\left[\mathbb H_\text{D}+\mathbb H_\text{S}\right]$ of equation~(\ref{BdG}) for a given pairing value $\Delta=0.3$, depicting in color only where the lowest singular value vanishes, $d_0=0$.

%-------------------ETA REGIMES--------------------------------------------
\begin{figure}[t!]
\centering
\resizebox{8.6cm}{!}{\input{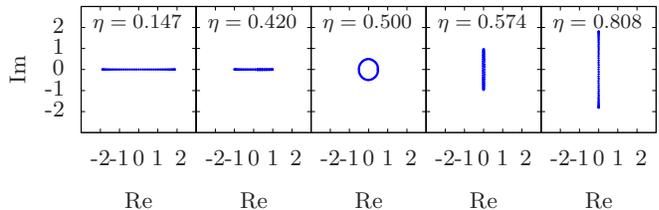}}
\caption{\label{Sequence}The coupling matrix spectrum goes from real ($\eta=0$)  to pure imaginary ($\eta=1$) values, showing the transition regime from the weak coupling (free-electrons with reduced mobility) to strong coupling where the pairing gives rise to Majorana quasiparticles.
  % $\eta_{1 \text{(Free-Top)}}$  $\eta_{2 \text{(Top-Strong)}}$
}
\end{figure}
%-------------------------------------------------------------------------
The parameter space of the Hamiltonian can be reduced by introducing the following parameterization $t=E_0 \cos^2( \pi \eta /2)$, $\Delta=E_0 \sin^2 (\pi \eta /2)$ with $\tilde H=\hat H/E_0$ and $\tilde \mu=\mu/E_0$  with $E_0=t+\Delta$ as an energy scaling where $\eta\in[0,1]$.  This parameterization determines four interesting regions which are shown in Fig.~\ref{Sequence}: $\eta=0$ for free electrons and real eigenvalues, $0<\eta\ll1$ or weak coupling where perturbation theory holds keeping the eigenvalues real, $\eta=1$ where the physics is dominated only by the pairing with purely imaginary eigenvalues, % $0<1-\eta\ll1$ for which a strong coupling expansion could be performed and the eigenvalues are still imaginary
and finally, the remaining values of $\eta$ form the fully dimerized region with complex eigenvalues and presenting unpaired Majoranas. Henceforth, we concentrate in the parameter region displaying Majoranas, they are the weak coupling and the fully dimerized regimes. 
%---------WEAK COUPLING COMPARISSON NUMERICS-PERTURBATION THEORY--
\begin{figure}[!t]\centering
\resizebox{8.3cm}{!}{\input{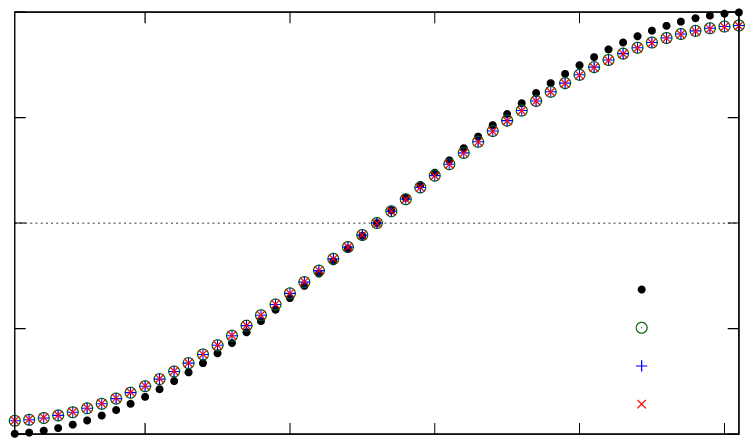}}
\caption{\label{Numerico-PT}Kitaev chain spectrum for $L=51$, $t=1$, $\mu=0$ and $\Delta=0.35$ where the perturbation theory is still valid.  As we can see, all the results coincide: Numerical diagonalization (${\color{red} \times}$), 3rd order perturbation theory (${\color{blue} +}$) and the fitting with an effective hopping (${\color{olive} \circ}$) from equation (\ref{effhopp}). The small dots in black ($\cdot$) represent the particular case of $\Delta=0$ as a guide, showing that the structure of the spectrum does not change with the small perturbation.}
\end{figure}
%----------------------------------------------------------------------
\section{Weak coupling regimen: $0\le \eta \ll 1$}\label{Perturbation}

In the Hamiltonian (\ref{BdG})  the $\mathbb H_\text{S}$ term can be treated as a small perturbation compared to the $\mathbb H_\text{D}$ term when $\eta$ is small enough. The correction to the energy for a given $\zeta$ up to third order is % presented in detail in appendix~\ref{PT-WC} and the final expression for the eigenvalue is
 given by,% (cf. appendix \ref{OSDFT}),
\begin{equation}
  E_\zeta^{(3)}=E_\zeta-\frac{8\Delta^2}{t(L+1)^2}\hspace{-0.4cm}\sum_{\tiny \begin{array}{c}\zeta'\\\zeta+\zeta'\text{odd}\end{array}}\hspace{-0.2cm}\frac{\textrm{S}_\zeta^2 \textrm{S}^2_{\zeta'}}{\left[\textrm{C}_\zeta-\textrm{C}_{\zeta'}\right]^3}.
\end{equation}

By analyzing this expression one can observe that, for $L\gg 1$, the perturbation only modify the band-width of the dispersion relation through an effective hopping parameter, $ t_\text{eff}=t-\frac{2\Delta^2}{t}$, and thus
\begin{equation}
E_\zeta^{(3)}\simeq-\mu-2 t_\text{eff}\cos\left(\frac{\pi\zeta}{L+1}\right).\label{effhopp}
\end{equation}
Figure \ref{Numerico-PT} presents the Kitaev spectrum for a given parameter set and a system size $L=51$. It shows the comparison between the full diagonalization of the Bogoliubov-de Gennes coupling matrix, the perturbation theory and the effective dispersion relation of Eq.~(\ref{effhopp}). We can see there the perfect agreement among all three curves. The free-particles spectrum ($\Delta=0$) is shown as a guide to the eyes making evident that the functional form of the spectrum does not change where the perturbation theory is still valid.

%---------PHASE DIAGRAM Parameterization-------------------------------
\begin{figure}[!ht]
\resizebox{8.3cm}{!}{\input{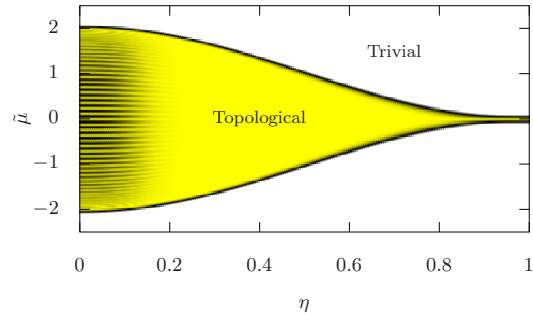}}  
\caption{\label{phase-diagram}(Color online.) Parametric $-\eta\tilde\mu$ phase space for the Kitaev chain. The topological phase is located in the color region bounded by the $ \tilde \mu = \pm  2 \left[\cos(\pi \eta / 2)\right]^2 $ curves, outside it the system remains in the trivial phase. At the left border one can see the wrinkles due to the finite size effect, for this plot we consider $L=51$.} 
\end{figure}
%-----------------------------------------------------------------------

\section{Fully dimerized limit: $0< \eta < 1$} \label{Wavefunction}

Let us consider the fully dimerized limit where zero energy modes (unpaired Majorana fermions) can be hosted embedded in a fermionic environment. For odd system sizes there is always a real eigenvalue and under the adequate parameter set it can be tuned to zero. 
In order to generate a $- \eta \tilde \mu$ phase diagram, we plot solely the minimum singular value , $d_0$, from equation~(\ref{BdG}).
%a singular value decomposition (SVD) to the parametric BdG matrix from equation~(\ref{BdG}), has been performed and solely the minimum singular value, $d_0$, has been plotted. 
In this fashion, we retrieve Fig.~\ref{phase-diagram} which, for $L=51$ sites, displays two distinct regions: the outside region in white where $d_0>0$, representing a trivial phase and the color region where the minimum singular value strictly annulates, $d_0=0$, and thus there should be a zero eigenvalue of the system.
Moreover, for the regime $\eta\ll1$, the system energy vanishes ($E_\zeta^{(3)}=0$) when the transcendental equation $\cos\left(\zeta\pi/(L+1)\right)=\tilde \mu/(-2t_\text{eff})$  holds. Only for certain values of $\tilde\mu$ there are zero-energy modes, hence Fig.~\ref{phase-diagram} presents wrinkles in the weak coupling regime. Nevertheless, this region smooths as the system size grows and the whole region homogenizes.

%-------------------MAJORANA'S WAVEFUNCTIONS---------------------------------
\begin{figure}[!hb]\centering
\resizebox{8.3cm}{!}{\input{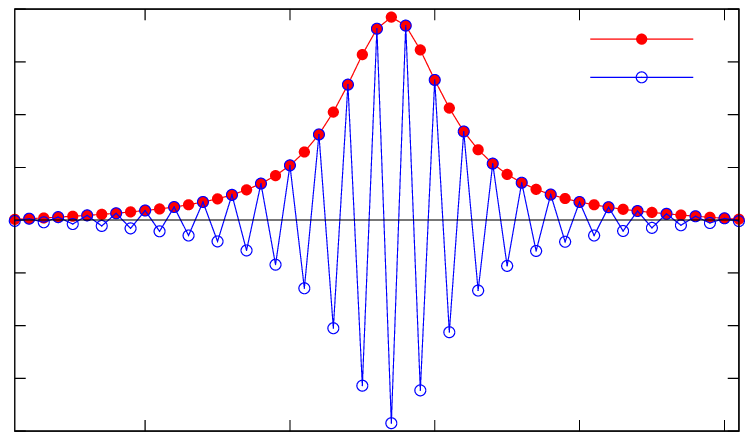}}\\\resizebox{8.3cm}{!}{\input{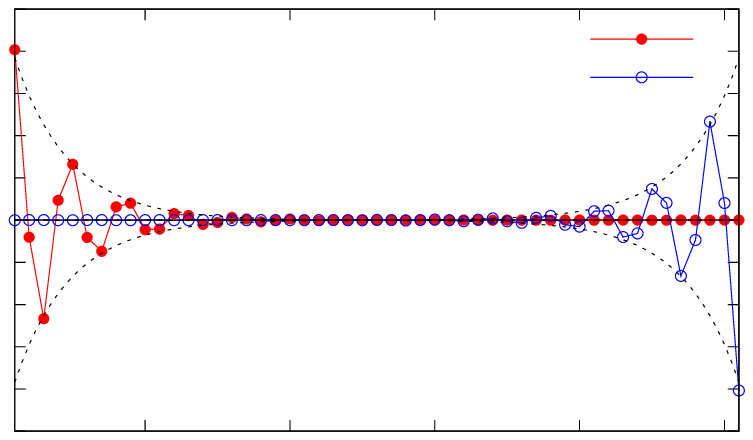}}\\
\caption{\label{ES_ZL} Zero energy eigenstates in (a) momentum and (b) position for both Majoranas kind A (blue) and B (red). The exponential fit is also presented in (b) showing the expected behavior for the edge states. The parameters considered are $L=51$, $\eta=0.3$ and $\mu=0.1$.}
\end{figure}
%---------------------------------------------------------------------------

In order to retrieve the zero-energy eigenstates, we consider the fact that wherever the coupling matrix $\mathbb H_\text{D}+\mathbb H_\text{S}$ has a zero eigenvalue, it also has a null determinant. In other words, its columns (rows) are linearly dependent vectors and thus it is possible to find at least one vector with dimension $L$ linearly independent to them. In this way, we take a random vector and diminish it by the projections upon each of the the columns (rows) of the interaction matrix. The result is a vector such that it is a right (left) eigenvector with an eigenvalue equals to zero. Figure~\ref{ES_ZL} shows the eigenfunctions of the zero-energy modes, Fig.~\ref{ES_ZL}(a) presents it in momentum space, $\phi_\zeta$, for both Majoranas species A and B. One can observed a delocalization for both wavefunctions being an oscillatory type for B-kind and its envelope given by A-kind. On the other hand, Fig.~\ref{ES_ZL}(b) displays both eigenfunctions in position space, $\phi_l$, a strong localization at the edges is clearly seen with exponential decays, distinct signatures of unbound Majorana states.

%\Ka{aqu\'i a\'un falta explicar la manera como se obtuvo los autoestados de los modos de energ\'ia cero!}

\section{Summary and remarks}\label{Conclusions}
We focus on the study of zero-energy modes of  the Kitaev  model putting special attention on a suitable transformation to bring the system from the position to the momentum representation considering the proper boundary conditions. The necessity lies in the fundamental fact that these are edge modes meaning the precondition of a finite chain. Hence, the customary Fourier transform, which holds solely for periodic boundary conditions, precludes the support of such unbound Majorana edge states. Therefore, in the present work, we propose the sine discrete Fourier transform type I to reach the momentum representation when using finite chains. %At this point, it is worth to point out that this sine transform is highly conveniet for any finite system, being rather correct and general, we use the Kitaev model as an instance.
Once the model is represented by the Majorana momentum operators, we use an extended Bogoliubov-de Gennes formalism and, by means of a SVD performed onto the resultant coupling matrix, the zero singular values are obtained, which in turn reflect the presence of zero eigenvalues. With that in hand, we obtain a parametric phase diagram of the Kitaev chain where we can distinguish the presence of the localized edge states. Since the coupling matrix is singular the usual diagonalization techniques involving matrix inversions are not appropriate, hence, a refined method to reach the corresponding zero-energy eigenvectors has been introduced and the eigenfunctions are also presented.

The different interwoven methodologies allow us to reduce the matrix dimension from an exponentially large dimension in the system size of $2^L\times 2^L$ to a linear dimension $L\times L$ for which a much larger number of sites may be studied in a finite chain. Additionally, we have learned how to treat the particular case of hard-wall boundary conditions and therefore we can apply the sine-transform to many systems where the finite size may enhance novel phenomena.% \Ka{aqu\'i la idea es recalcar que la transformada funciona para cualquier tipo de cadena finita!!!, ese creo yo debería ser el final remark.}
\begin{acknowledgments}
The authors acknowledge the support from CIBioFi and the Colombian Science, Technology and Innovation Fundation--COLCIENCIAS ``Francisco Jos\'e de Caldas'' under project 1106-712-49884 (contract No.264-2016) and --General Royalties System (Fondo CTeI-SGR) under contract No. BPIN 2013000100007. 
\end{acknowledgments}

\appendix
\section{Orthonormalized sine discrete Fourier transform type I }\label{OSDFT} 
  The basis orthonormality is given by 
  \begin{align}
    &\langle k_\zeta^-|k_{\zeta'}^-\rangle\notag\\
    =&\frac{2}{L+1}\sum_{l=1}^{L}\sum_{l'=1}^{L}\sin\left(\frac{\pi}{L+1}\zeta l\right)\sin\left(\frac{\pi}{L+1}\zeta'l'\right)\expectation{\varnothing}{\hat a_l\hat a_{l'}^\dagger}{\varnothing},\notag\\
    =&\frac{1}{L+1}\sum_{l=1}^{L}\Bigg[\frac{1}{2}\{1+(-1)^{\zeta-\zeta'}\}-1\notag\\
      &-\left(\frac{1}{2}\{1+(-1)^{\zeta+\zeta'}\}-1\right)\Bigg]\notag\\
    =&\frac{1}{L+1}\sum_{l=1}^{L}\frac{1}{2}\underbrace{\Bigg[(-1)^{\zeta-\zeta'}-(-1)^{\zeta+\zeta'}\Bigg]}_{=0, ~\text{if } \zeta \ne \zeta'},
  \end{align}
  where we have used %the trigonometric identity $\sin(A)\sin(B)=\frac{1}{2}\left(\cos(A-B)-\cos(A+B)\right)$ with $A-B=\frac{\pi}{L+1}(\zeta-\zeta')l$ and
  the summation 
  \begin{equation}
\sum_{p=1}^q\cos{px}=\frac{1}{2}\left(1+\frac{\sin\left((q+\frac{1}{2})x\right)}{\sin\left(\frac{x}{2}\right)}\right)-1.
  \end{equation}
  In the remaining case, $\zeta=\zeta'$, we have
  \begin{equation}
    \langle k_\zeta^-|k_\zeta^-\rangle=\frac{2}{L+1}\sum_{l=1}^{L}\sin^2\left(\frac{\pi}{L+1}\zeta l\right)=1,
  \end{equation}
  here we have used the sum
  \begin{equation}
    \sum_{p=1}^q\sin^2(px)=\frac{q}{2}-\frac{\cos((q+1)x)\sin(qx)}{2\sin(x)}.
    \end{equation}
  Hence the orthornormalization, $\langle k_\zeta^-|k_{\zeta'}^-\rangle=\delta_{\zeta,\zeta'}$, is granted for all basis states.
  Finally, the completeness relation is given by,
  \begin{equation}
\sum_{l=1}^{L}\sin\left(\frac{\pi l}{L+1}\zeta\right)\sin\left(\frac{\pi l}{L+1}\zeta'\right)=\frac{L+1}{2}\delta_{\zeta,\zeta'}.
  \end{equation}

  Having the sine transform, we can now determine how to relate the position and momentum space operators using it as shown above in equation (\ref{sine-transform}).

%  \begin{align}
%\hat a_l=\sqrt{\frac{2}{L+1}}\sum_{\zeta=1}^{L}\sin\left(\frac{\pi l \zeta}{L+1}\right)\hat a_\zeta.
%  \end{align}

\bibliography{biblioMajoranas}

\end{document}